\documentstyle[aps,prb,epsfig,twocolumn]{revtex}

\begin{document}

\bibliographystyle{prsty}

\draft

\wideabs{
\title{ Stochastic Force Defined Evolution in Dynamical Systems }
\author{ Ping Ao }
\address{ Institute for Systems Biology, 1441 N. 34 St., Seattle, WA 98103 }
\address{Department of Physics, University of Washington, Box 351560, Seattle,
WA 98195 }
\address{ }
\date{ submitted to Physical Review Letters on Nov. 21, 2002 }

\maketitle

\begin{abstract}
Dynamics near and far away from thermal equilibrium is studied within the
framework of Langevin equations. A stochasticity-dissipation relation is
proposed to emphasize the equal importance of the stochastic and deterministic
forces in describing the system's evolution and destination. It is a
generalization of the fluctuation-dissipation theorem. Close to a stationary
state an explicit construction of the potential energy is given through a gauged
singular decomposition. Possible directions to extend the present study into
generic situations are pointed out.
\end{abstract}

\pacs{PACS numbers: 87.23.Kg,05.70.Ln,47.70.-n, 87.23.Ge }
}


In biology, chemistry, and physics, a system under investigation can often be
well modeled by a set of deterministic differential equations.\cite{kaplan}
Those deterministic differential equations, describing phenomena on a
macroscopic level ranging from near to far from thermal equilibrium, are
usually motivated by empirical observations. In addition, it is frequently
found that adding the noise or stochastic force into them can result in a
better description of systems, or, enables investigators to get certain
desired results, such as particular stationary distributions. This is usually
done in an {\it ad hoc} manner. The important question arises that whether or
not the noise is an integral part of the mathematical modeling, such that the
stochastic force has a transparent role and is directly connected to
experimental observations. Ample experimental evidence from physics to
biology, particularly the studies of the properties near thermal equilibrium in
chemistry and physics\cite{miller} and of the evolution of species in
biology\cite{delbruck}, suggests the affirmative answer. In attempt to address
this question, our first response may be to resort to a microscopic derivation
under appropriate conditions. This would certainly be advantageous, because it
shows the coherence between the macro and micro modeling. However, often in
reality, apart from a heuristic justification, not only the microscopic
derivation is completely missing in a reductionist's point of view, the very
existence of the corresponding microscopic base can be called into doubt. Even
in a situation that such a microscopic base exists, as typically believed in
physics, the derivation involves a careful consideration of the emerging
properties and can be highly nontrivial. Therefore, granted the autonomy of the
macroscopic description, can we formulate the deterministic equations and
the noise in a coherent manner? In this letter we will present an explicit
construction leading to the positive answer. The presentation is inspired by
the methodology of Onsager \cite{onsager} in his classical investigation of
dynamics near thermal equilibrium. We have generalized Onsager`s results in two
important directions: We elevate the stochastic force to the position of equal
importance to that of the deterministic equation, and explicitly construct the
stationary distribution for the state variable in typical situations far away from thermal
equilibrium  where the detailed balance condition does not hold.

To avoid the unnecessary mathematical complication, we will restrict the
differential equations to be a set of first order ordinary differential
equations. This corresponds to the homogeneous and well-stirred situation in a
chemical reaction network widely encountered in biology and chemistry. Let the
$d$ component state variable be presented by $ x^{\tau} = (x_1, ... , x_d )$, a
vector in $d$ dimensional real space, the deterministic dynamical equation is 
\begin{equation} 
   \dot{x} = f(x) \; .
\end{equation}
Here the superscript $\tau$ denotes the transpose, the dot over the state
variable denotes the derivative with respect to time $t$, $\dot{x} = {d x}/ {dt
}$, and the $n$ component force $f^{\tau}(x) = ( f_1(x), ... , f_d(x) )$ with
$f_i(x) = f_i(x_1, ... , x_d ), \; i=1, ... , d$ is a smooth function
specifying the evolution of the state variable in a deterministic manner.
Eq.(1) is a shorthand writing for ${d x_i}/{d t} = f_i(x), \; i=1, ... , d$.The force $f$ is specified by a set of controlling parameters which we will not further discuss in this letter, except we assume the force is explicitly time independent. The state variable can be the particle position in physics, or the concentrations of reactants in chemistry, or the protein expression levels in biology. There are two important features differ Eg.(1) from the usual Hamiltonian dynamics. First, the divergence of the force is finite except for a few isolated points: $ \partial \cdot f \neq 0 $, where $\partial \cdot f = \sum_{i=1}^{d} \partial_i f_i(x)$, and $ {\partial}_i = {\partial } / {\partial x_i }$. The finite divergence implies that both the energy and the volume of phase space are not conserved. Second, the skew matrix $\partial \times f$ generated by the skew operation $ (\partial \times f)_{ij} = \partial_i f_j(x) - \partial_j f_i(x) $ is in general finite, too:
$ \partial \times f \neq 0 $.
In two and three dimensions, $d = 2,3$, the skew operation reduces to the usual
curl operation, $\nabla \times $, and the skew matrix can be represented by a
vector. The finiteness of the skew matrix is a characteristic of the system far
away from thermal equilibrium. It is believed that it can cause the absence of
a potential\cite{nicolis}. We will, however, present an explicitly construction
of a potential function. 

To emphasize the essential idea of our method, we further simplify our problem
by considering the dynamics near a stable point specified by zero force, $f =
0$. We will choose this stable point to be the origin in our $d$ dimension
state space. A stable point represents a stable stationary state. Its existence
is evident in chemistry and physics. In biology, where systems are believed to
operate under the condition far away from thermal equilibrium, the existence of
such a stable stationary state has also been demonstrated.\cite{baikai}. Hence,
to the linear order in $x$, the leading order close to the stable point $x=0$,
$f = - F x$ or $f_i = - \sum_{j=1}^{d} F_{ij} x_j$. Adding the noise, the
stochastic force, into Eq.(1), we have
\begin{equation}
   \dot{x} = - F x + \zeta (t) \; ,
\end{equation}
where the stochastic force $\zeta^{\tau}(t) = (\zeta_1(t), ... , \zeta_d(t))$,
which is chosen to be represented by a Gaussian and white noise, with zero
mean, $ < \zeta(t) > = 0$, and the variance 
\begin{equation} 
  < \zeta(t) \zeta^{\tau}(t') > = 2 D \delta (t-t') \; .
\end{equation}
Here $\delta(t)$ is the Dirac delta function and the constant diffusion matrix
$D$ is explicitly defined by $ < \zeta_i(t) \zeta_j(t') > = 2 D_{ij} \delta
(t-t')$. The average $< ... >$ is carried out in the distribution function for
the noise, not in terms of distribution function for the state variable. Unless
explicitly specified, whenever it can be defined, the temperature in the
present letter is always set to be 1: $k_B T = 1$. This is equivalent to a
rescale of energy. To ensure the independence of all components of the state
variable near the stable point, we will require the determinant of the constant
force matrix $F$ to be finite: $\det (F) \neq 0$. We will call Eq.(2) the
standard Langevin equation, or, the standard form of the stochastic
differential equation.\cite{vankampen,risken,gardiner} 

Now, we assert that there exists a transformation, such that we can transform
Eq.(2) into the following form:
\begin{equation}
  ( S + A ) \dot{x} = - \partial u(x) + \xi(t) \; , 
\end{equation}
with following properties: $S$ is a constant symmetric and semi-positive
matrix, $A$ a constant antisymmetric matrix, and $u(x)$ a single valued scale
function to be specified below. The stochastic force $\xi^{\tau}(t)=(\xi_1(t),
... , \xi_d(t)) $ is Gaussian and white and has the same origin as that of
$\zeta(t)$ . We further require that the determinant of $S + A$ is finite,
$\det(A+S) \neq 0$, for the same reason of requiring $\det(F) \neq 0$. We will
call such a transformation from Eq.(2) to (4) the singular decomposition.
Clearly, the symmetric matrix $S$ plays the role of friction, and the
antisymmetric matrix $A$ the role of `magnetic' field: `Energy' dissipation is
always non-negative, $\dot{x}^{\tau} S \dot{x} \geq 0 $, and the transverse
force, a `Lorentz' force, does no work, $\dot{x}^{\tau} A \dot{x} = 0 $. The
function $u(x)$ hence acquires the meaning of potential energy. While given
Eq.(4) a unique Eq.(2) can be obtained, we point out that if the stochastic
forces would be ignored, that is, $\zeta(t) = \xi(t) = 0$, the singular
decomposition would not be unique. There would exist a family of singular
decompositions transforming Eq.(2) to (4). One would be easily able to verify
this observation by multiplying an arbitrary finite constant to Eq.(4): a large
friction and a large potential has the same deterministic dynamics of a small
friction and a small potential. Thus for the same deterministic dynamics
specified by Eq.(1) ( $\zeta(t) = 0$ in Eq.(2) ), the underlying system
properties such as the energy dissipation determined specified by Eq.(4) with
$\xi(t) = 0$ would be totally different. This observation strongly indicates
that the stochastic force is an integral part of the system's dynamics: the
stochastic force introduces an intrinsic scale into the problem. Specifically,
requiring the uniqueness of the singular decomposition, we impose the following
condition which links the stochastic force to the dissipation matrix: 
\begin{equation}
    < \xi(t) \xi^{\tau}(t') > = 2 S \delta(t-t') \; .
\end{equation}
Together with $< \xi(t) > = 0$, Eq. (5) will be called the
stochasticity-dissipation relation, and Eq.(4) will be called the normal
Langevin equation, or, the normal form of the stochastic differential
equation. The singular decomposition under this stochasticity-dissipation
relation will be called the gauged singular decomposition.

By our construction, the same dynamics is described by either the standard
Langevin equation, Eq.(2), or, the normal Langevin equation, Eq.(4).
Eliminating $\dot{x}$ from those two equations leads to
\begin{equation}
   (S + A ) [ - F x + \zeta(t) ] = - \partial u(x) + \xi(t) \; .
\end{equation}
Because the stochastic forces come from the same source, and their dynamics is
independent of that of the state variable, separately we have 
$ ( S + A ) F x = U x $,
where we have set $\partial u(x) = U x$, and 
$ (S + A ) \zeta(t) = \xi(t) $.
Since $x$ can be an arbitrary state vector, we have
\begin{equation}
   (S + A ) F = U \; .
\end{equation}
Using Eq.(5) and $< (S + A) \zeta(t) \zeta^{\tau}(t') (S - A) > = < \xi(t)
\xi^{\tau}(t') >$, we obtain the following generalized Einstein relation
between the diffusion matrix $D$ and the friction matrix $S$:
\begin{equation}
  (S + A) D ( S - A ) = S \; .
\end{equation}
Eq.(8) suggests a duality between the standard and the normal Langevin
equations: large friction matrix implies small diffusion matrix, and vice
versa.

We next prove the existence and uniqueness of the gauged singular decomposition
by an explicit construction. Using the fact that the potential matrix $U$ is
symmetric as required by $\partial \times \partial u(x) = 0$, we have
\begin{equation}
   (S + A ) F - F^{\tau} (S - A ) = 0 \; .
\end{equation}
Defining an auxiliary matrix $G \equiv (S + A )^{-1}$, the generalized
Einstein relation, Eq.(8), and the above equation lead to the following two
coupled inhomogeneous linear equations for $G$ and its transpose:
 \begin{equation}
   G F^{\tau} - F G^{\tau} = 0 \; , 
\end{equation}
and 
\begin{equation}
   G + G^{\tau} = 2 D \; .
\end{equation}
The symmetric part of the auxiliary matrix $G$ is readily available from Eq.(11). Equations similar to Eq.(11) have been discussed
Before in the context of Ornstein-Uhlenbeck process\cite{vankampen,risken,gardiner,lebowitz} The antisymmetric part, $Q \equiv (G - G^{\tau} )/2$, can be formally expressed as a series after a straightforward matrix manipulation:
\begin{equation}
  Q = \sum_{n=1}^{\infty}(-1)^{n}[ F^{-n} D (F^{\tau} )^{n} 
         - F^{n} D (F^{\tau} )^{-n}  ]  \; . 
\end{equation}
Having obtained the auxiliary matrix $G$ in terms of the force matrix $F$ and
the diffusion matrix $D$, $G = D + Q$, the gauged singular decomposition is
uniquely determined: 
\begin{equation}
  \left\{ \begin{array}{lll}
  U & = & G^{-1} F \\
  S & = & [ G^{-1} + (G^{\tau} )^{-1} ] /2 \\
  A & = & [ G^{-1} - (G^{\tau} )^{-1} ] /2
  \end{array} 
     \right. \; . 
\end{equation} 
The potential energy is $u(x) = x^{\tau} U x /2$. This completes our proof of the existence and uniqueness of the gauged singular decomposition.

With the energy function in the present problem as the potential energy
$ u(x) $, the stationary distribution function $\rho_0(x)$ for the state variable should be given by 
\begin{equation}
   \rho_0(x) = \frac{1}{Z} \; \exp\{ - u(x) \} \; ,
\end{equation}
with the partition function $Z = \int d^d x \; \exp\{ - u(x) \} $. The normal
Langevin equation, Eg.(4), allows a particular easy identification of the energy
function. Without the stochastic force, no energy can be uniquely defined from Eq.(1) or
Eq.(2), and the singular decomposition is not unique, too. The available volume of phase space for the state variable would be shrank to zero, to a point defined by the stationary state.

From either the standard or the normal Langevin equations, it is
straightforward to obtain the corresponding Fokker-Planck equation for the
distribution function $P(x,t)$ of state variable\cite{risken}:
\begin{equation}
  \partial_t P(x,t) + \partial \cdot [ (f - D \partial ) P(x,t) ] = 0 \; .
\end{equation}
Here $\partial_t = {\partial }/{\partial t}$. If the probability current
density is defined as $J(x,t) \equiv ( f - D \partial ) P(x,t)$, the
Fokker-Planck equation is a statement of the probability continuity: $
\partial_t P(x,t) + \partial \cdot J(x,t) = 0 $. The stationary state corresponds to the condition 
\begin{equation}
   \partial \cdot J_0(x) = 0 \; .
\end{equation}
Particularly, the condition
\begin{equation}
   J_0(x) = 0
\end{equation}
has been called the detailed balance condition.\cite{risken,vankampen,gardiner}
One can check that the stationary distribution $P_0(x)$ in Eq.(14) is indeed the time independent solution of the Fokker-Planck equation: The stationary probability current $J_0(x) = - G A G^T U x \; P_0(x) $ and $\partial \cdot J_0(x) = 0 $. Unless the transverse matrix $A$ is zero, the detailed balance condition does not hold. This suggests a natural description of a stationary cyclic motion by the normal Langevin equation.

One comment is in order: The uniqueness of the gauged singular decomposition is guaranteed by the conditions $\det(F) \neq 0$ and $\det( S + A ) \neq 0$, which leads to the Gaussian distribution for the stationary distribution for state variable. Hence near thermal equilibrium the stochasticity-dissipation relation is equivalent to the fluctuation-dissipation theorem.\cite{onsager,miller}
Nevertheless, in the normal Langevin equation no assumption is made on the underlying thermal dynamics. The real difference is that by the
stochasticity-dissipation relation the emphasis is on the stochastic force, therefore on dynamics, and by the usual fluctuation-dissipation theorem the emphasis is on the stationary state distribution, therefore on statics. In this sense one may regard the stochasticity-dissipation relation as a generalization of the fluctuation-dissipation theorem. 

To further illustrate the coherence and generality of the present method, we
discuss three aspects of the normal Langevin equation and the gauged singular
decomposition. We first look further into the detailed balance condition. In
order to have the stationary probability current $ J_0(x) = 0$, the necessary and
sufficient condition is the transverse matrix to be zero: $A =0$. This is
equivalent, according to Eq.(13), $G = G^{\tau}$. It follows that both the matrix $U
F$ and matrix $ S F $ are symmetric. For a symmetric force matrix $F$, this
further leads to the condition $ S F = F S $, that is, $S$ and $F$ commute.
This implies that, even for a symmetric force matrix $F$, not every choice of
the friction matrix $S$, therefore the potential matrix $U$, can guarantee the
detailed balance condition.

Second, we consider four examples from four different fields, where either the
force matrix $F$ is explicit asymmetric or the transverse matrix $A$ is finite.
They belong to two special but important classes where the direct connection
between the deterministic force and the stochastic force can be established
microscopically. The first example is well-known in physics: the charged
particle moving in a magnetic field, the magneto-transport in solid state
physics.\cite{ziman} The second example is the emergent vortex dynamics in
superfluids and superconductors, whose microscopic derivation is very
technical.\cite{at} When the damping is strong enough, the equation of motions
are already in the form of normal Langevin equation. For both cases at two
dimension, $d=2$, we can identify the friction and transverse matrices as:
\begin{equation}
  S + A = \left( \begin{array}{cc} 
                                   \eta & b \\
                                   - b & \eta 
                         \end{array} \right)
\end{equation}
In both cases the friction and stochastic force can be formulated as the
results of the coupling between the system and a reservoir. The stochastic
dynamics can be formulated within the Hamiltonian framework cherished in
physics.\cite{leggett} 
Dynamical equations already in the form of standard Langevin equation are the
Lotka-Volterra equation for species competition \cite{kaplan}, and the toggle
equation for the stability of gene switch\cite{gardner}, our third and fourth
examples. In the corresponding force matrix $ F $ 
\begin{equation}
    F_{12} \neq F_{21} \; .
\end{equation} 
The Lotka-Volterra and toggle equations describe processes belong to the
generic predator-prey or growth-decay process, where the diffusion matrix $D$ can
be obtained based on the knowledge of the deterministic equation when large
number of birth and death events occur on the macroscopic time
scale.\cite{mcquarrie,vankampen,gardiner}
  
In most cases, however, both the intrinsic and the extrinsic noise coexist. They are
equally important and can be determined experimentally.\cite{elowitz} The
stochasticity-dissipation relation treats them on the equal footing to
determine the gauged singular decomposition. Quantitative and global
predictions can be made based on the normal Langevin equation.

Finally, we consider the classical example of a damped harmonic particle in one
dimension to illustrate the consistency of our method. The celebrated equation
of motion for momentum $p$ and coordinate $q$
is\cite{vankampen,risken,gardiner}
\begin{equation}
  \left\{ \begin{array}{ccl}
    \dot{q} & = & {p}/{m} \\
    \dot{p} & = & - {\eta}/{m} \; p - k q + \zeta(t) 
   \end{array} \right.
\end{equation}
Here $ < \zeta(t) > = 0$ and $ < \zeta(t) \zeta(t')> = 2 \eta \delta (t-t') $, 
$m$ is the mass of the particle, and $k$ the spring constant. The equilibrium
distribution function is\cite{vankampen,risken,gardiner} $P_0(x) = 1/Z \;
\exp\{ - ( p^2 /2m + k q^2 /2 ) \}$, because the total energy can be readily
identified as $ E = p^2 /2m + k q^2 /2$, the sum of kinetic energy and
potential energy. 
Let $x_1 = q, x_2 = p$, the force matrix $F$ and the diffusion matrix $D$ can
be easily identified as 
\begin{equation} 
  F = \left( \begin{array}{cc} 
                                   0 & - {1}/{m} \\
                                   {k} & {\eta}/{m} 
                         \end{array} \right) \; , \;
  D = \left( \begin{array}{cc} 
                                   0 & 0 \\
                                   0 & \eta 
                         \end{array} \right) \; .
\end{equation}
By a straightforward calculation, the gauged singular decomposition consisting
of the friction matrix $S$, the transverse matrix $A$ and the potential matrix
$U$ is found as,
\begin{equation} 
  U = \left( \begin{array}{cc} 
              k & 0 \\
              0 & {1}/{m} 
      \end{array} \right) , 
  S = \left( \begin{array}{cc} 
              \eta & 0 \\
              0 & 0 
             \end{array} \right) , 
  A = \left( \begin{array}{cc} 
                 0 & 1 \\
               -1 & 0 
             \end{array} \right) \; ,
\end{equation}
which gives the same energy function, as it should.

In conclusion, we have explicitly constructed the normal Langevin equation from
the standard Langevin equation via a gauged singular decomposition. The
fluctuation-dissipation theorem is generalized to the stochasticity-dissipation
relation to emphasize the dynamical nature of the stochastic force. We have
obtained the potential energy which determines the stationary distribution even
when the detailed balance condition does not hold. Finally, we point out that the
normal Langevin equation enables the generalization in several directions. 
It is straightforward to generalize  the approach to unstable fix points.  
Taking corresponding limits of the dimension of the phase space goes to infinite, 
t may be applied to the stochastic partial differential equations. 
We note that it is already in the form ready for the color noise case as demonstrated 
in the study of the dissipative dynamics; and if we view the present construction 
as a local approximation, the extension to the generic nonlinear situation is implied.

Discussions with Lee Hood, David Thouless, Lan Yin, and Xiaomei Zhu 
are highly appreciated. This work was supported in part 
by Institute for Systems Biology.


\begin{thebibliography}{99}

\bibitem{kaplan}
   D. Kaplan and L. Glass, Understanding nonlinear dynamics, Springer-Verlag,
   Berlin, 1995.
\bibitem{miller}
   D.G. Miller, in Foundations of continuum thermodynamics, edited by J.J.D. Domingos,  
   M.N.R. Nina,  and J.H. Whitelaw, John Wiley and sons, New York, 1973. 
\bibitem{delbruck}
  S.E. Luria and M. Delbruck, Genetics, 28 (1943) 491; 
  L.H. Hartwell {\it et al.}, Genetics: from genes to genomes, McGraw-Hill,
  Boston, 2000.
\bibitem{onsager}
   L. Onsager, Phys. Rev., 37 (1931) 405; {\it ibid}, 38 (1931) 2265;
   L. Onsager and S. Matchlup, {\it ibid} 91 (1953) 1505; 
   S. Machlup and L. Onsager, {\it ibid}, 91 (1953) 1512. 
\bibitem{nicolis}
   G. Nicolis and I. Prigogine, Self-organization in nonequilibrium systems:
    from dissipative structure to order through fluctuations, John Wiley and
    sons, New York, 1977.
\bibitem{baikai}
   U. Alon {\it et al.}, Nature 397 (1999) 168. 
\bibitem{vankampen}
   N.G. van Kampen, Stochastic processes in physics and chemistry, Elsevier,
   Amsterdam, 1992.
\bibitem{risken}
   H. Risken, The Fokker-Planck equation, Springer, Berlin, 1989.
\bibitem{gardiner}
   G.W. Gardiner, Handbook of stochastic methods for physics, chemistry and the
    natural sciences, Springer-Verlag, Berlin, 1983.
\bibitem{lebowitz}
   G.L. Eyink, J.L. Lebowitz, and H. Spohn, J. Stat. Phys. 83 (1996) 385.
\bibitem{ziman}
  J.M. Ziman, Electrons and phonons; the theory of transport phenomena in
  solids, Clarendon Press, Oxford, 1962. 
\bibitem{at}
   P. Ao and D.J. Thouless, Phys. Rev. Lett.70 (1993) 2158; P. Ao and X.-M.
   Zhu, Phys. Rev. B, 60 (1999) 6850. 
\bibitem{leggett}
   A.J. Leggett, in Quantum tunnelling in condensed media, edited by Yu. Kagan
   and A.J. Leggett, North-Holland, Amsterdam, 1992
\bibitem{gardner}
   T.S. Gardner {\it et al.}, Nature 403 (2000) 339.
\bibitem{mcquarrie}
   D.A. McQuarrie, J. Appl. Prob. 4 (1967) 413.
\bibitem{elowitz}
   M.B. Elowitz {\it et al.}, Science 297 (2002) 1183.

\end{thebibliography}
\end{document}